\documentclass{aa}
\usepackage{graphicx}
\usepackage{epsfig}
\usepackage{graphics}
\newcommand{\be}{\begin{equation}}
\newcommand{\ee}{\end{equation}}

\begin{document}
\title{Runaway massive stars as variable gamma-ray sources}

   \author{M.~V. del Valle\inst{1,2,}\thanks{Fellow of CONICET, Argentina}
           \and G.~E. Romero,\inst{1,2,}\thanks{Member of CONICET, Argentina}
           }

   \offprints{Mar\'{\i}a V. del Valle : \\ {\em maria@iar-conicet.gov.ar}}
   \titlerunning{Runaway massive stars as variable gamma-ray sources}

\authorrunning{del Valle \& Romero}  

\institute{Instituto Argentino de Radioastronom\'{\i}a, C.C.5, (1894) Villa Elisa, Buenos Aires,
Argentina. \and Facultad de Ciencias Astron\'omicas y Geof\'{\i}sicas,
Universidad Nacional de La Plata, Paseo del Bosque, 1900 La Plata, Argentina.}

\date{Received / Accepted}

% \abstract{}{}{}{}{}
% 5 {} token are mandatory

\abstract 
% context heatding (optional) % {} leave it empty if necessary 
{Runaway stars are ejected from their formation sites well within molecular cores in giant dark clouds. Eventually, these stars can travel through the molecular clouds, which are highly inhomogeneous. The powerful winds of massive runaway stars interact with the medium  forming bowshocks. Recent observations and theoretical modelling suggest that these bowshocks emit non-thermal radiation. As the massive stars move through the inhomogeneous  ambient gas the physical properties of the bowshocks are modified, producing changes in the non-thermal emission.} 
% aims heading (mandatory)
{We aim to compute the non-thermal radiation produced in the bowshocks of runaway massive stars  when travelling through a molecular cloud.} 
% methods heading (mandatory) 
{We calculate the non-thermal emission and absorption for two types of massive runaway stars, an O9I and an O4I, as they move through a density gradient. }
% results heading (mandatory)
{We present the spectral energy distributions for the runaway stars modelled. Additionally, we obtain light curves at different energy ranges. We find significant variations in the emission over timescales of $\sim$ 1 yr.}
% conclusions heading (optional), leave it empty if necessary 
{We conclude that bowshocks of massive runaway stars, under some assumptions, might be variable gamma-ray sources, with  variability timescales that depend on  the medium density profile. These objects might constitute a population of  galactic gamma-ray sources turning on and off within years.}


\maketitle

\section{Introduction}

Runaway massive stars have large spatial velocities ($v_{\star}>30$ km s$^{-1}$) (e.g., Gies \& Bolton 1986; Tetzlaff, Neuh\"{a}user \& Hohle 2011). They move supersonically through the interstellar medium (ISM) forming a  bowshock pointing in the direction of the star velocity (e.g., Van Buren, Noriega-Crespo \& Dgani 1995). The shocked material is heated by the stellar radiation field, and the swept dust re-emits at infrared (IR) wavelengths (e.g., Van Buren \& McCray 1988; Kobulnicky, Gilbert \& Kiminki 2010). 

The high proper velocities of runaway stars can be produced by the supernova explosion of a presumed binary companion (Blaauw 1961), or by dynamical ejection (Leonard \& Duncan 1988). Recently, Fujii \& Zwart (2011) argued that the velocity originates from strong gravitational interactions between single stars and binary systems in the centres of stellar clusters. Runaway massive stars are usually present around young star clusters. 

Bowshocks of early-type massive stars might produce non-thermal radiation (see del Valle \& Romero 2012), a fact supported by several observations. Radio non-thermal emission was detected from the bowshock of the runaway star BD+43${^{\circ}}$3654 (Benaglia et. al 2010). Recently, the bowshock  of the star AE Aurigae was detected at X-ray energies, and the emission is well described by a power-law spectrum that can be modelled as inverse Compton (IC) up-scattering of IR photons. (L\'opez-Santiago et al. 2012). Finally,  the bowshock of the well-known massive star HD 195519 has been associated with a {\it Fermi} source (see del Valle, Romero \& De Becker 2013). 

Runaway stars can eventually travel through the molecular cloud (MC) where they were formed, and interact with dense structures. The bowshock-medium interactions produce variable non-thermal emission as the star moves through the MC. In this work, we propose that runaway early-type stars  moving within  MCs can be variable gamma-ray sources. These objects might be  counterpart of some of the unidentified variable gamma-ray sources, concentrated towards the galactic plane. We adopt the model developed by del Valle \& Romero (2012) to compute spectral energy distributions and light curves for different types of stars. We also calculate the photon absorption along the whole spectrum.

In the next section, we briefly describe bowshocks of runaway stars. In Sec. \ref{mc},  we briefly discuss the  molecular cloud structure we adopted. In the next section, we present the non-thermal emission and absorption calculations we implemented for O4I and  O9I  stars. The main results are also given in this section. Finally, in Sec. \ref{end}, we discuss the variability of the emission and we offer our conclusions.

\section{Bowshocks of massive runaway stars}\label{bw}

Significant research has been done on bowshock modelling (e.g., Van Buren \& McCray 1988; Van Buren et al. 1990; Bandiera 1993; Brighenti \& D\'ercole 1995; Chen, Bandiera \& Wang 1996; Wilkin 1996; Comer\'on 1997; Chen \& Huang 1997; Comer\'on \& Kaper 1998; Wilkin 2000; Wareing, Zijlstra \& O'Brien 2007). The collision of a stellar wind, of mass-loss rate  $\dot{M}_{\rm w}$, density $\rho_{\rm w}$ and terminal velocity $V_{\rm w}$, with  the ISM, of density $\rho_{\rm a}$, around a runaway star results in a system of two shocks. The ram pressure of the wind and the ISM balances at some distance  from the star, i.e., $\rho_{\rm w}V_{\rm w}^{2} = \rho_{\rm a}V_{\star}^{2} $, where ${\rho}_{\rm w} = \dot{M}_{\rm w}/4{\pi}R^{2}V_{\rm w}$.  Here $R$ is the radial distance from the star. The value of $R$ where this occurs is defined  as the standoff radius $R_{0}$:

\begin{equation}
{R_{0} = \sqrt{\frac{ \dot{M}_{\rm w} V_{\rm w}}{4 \pi {\rho}_{\rm a} V_{\star}^{2}}}}
.
\label{R0}
\end{equation}

As can be seen from Eq. (\ref{R0}), the standoff radius decreases for a denser ambient medium. This is because the ISM ram pressure becomes stronger. 

 Bowshocks of runaway stars are imaged in the IR because of the emission produced by the heated gas and dust that they sweep. The heating of the material can be produced by the UV emission of the runaway star or with the radiation of the shocked gas in the post-shock region. A simple energetic analysis shows that the former dominates by at least an order of magnitude (e.g., Van Buren \& McCray 1988). The kinetic power of the stellar wind is $L{\rm w}$ $=$ $1/2\,\dot{M}_{\rm w}V_{\rm w}^{2}$ $\sim$ $3.14\times10^{35}\,(\dot{M}_{\rm w}/{\rm M_{\odot} yr^{-1}})(V_{\rm w}/{\rm km \, s^{-1}})^{2}$ erg s$^{-1}$. The available power for heating the gas is a fraction $\xi$ of the wind power, i.e., $\xi\,L_{\rm w}$. On the other hand, the power from the star luminosity is $3.84\times 10^{33}\,(L_{\star}/L_{\odot})$ erg s$^{-1}$. Again, the available power for heating the gas and dust with this radiation is a fraction of $L_{\star}$. The power shock/radiation ratio for the case of an O4I star is $L_{\rm w}$/$L_{\rm UV}$ $\sim$ $1.52\times10^{38}/2.69\times10^{39}$ $<\,1$ (see the stellar parameters in Table~\ref{tabla}; the factor $\xi$ is the same in both cases). It is clear then that the dominant heating mechanism is stellar radiation. In comparison, the radiation from the shocked gas plays a minor role.  

In the case of a concrete star, such as $\zeta$ Oph (a well-known runaway star, e.g., Peri et al. 2012), $L_{\rm w}$ $\sim$   $3.14\times10^{35}(10^{-7})(1.5\times10^{3})^{2}$ erg s$^{-1}$ $\sim$ $7\times 10^{34}$ erg s$^{-1}$. The observed IR luminosity from the bowshock reaches values of $5.5\times10^{35}$ erg s$^{-1}$, i.e., it is higher than even the mechanical power of the wind, showing that the stellar luminosity is the main heating source (e.g., Povich et al. 2008).

The characteristic temperature of dust emission, $T_{\rm IR}$, can be estimated  using a simplified dust model proposed by Draine \& Lee (1984). Given a radiation field and the grain absorption efficiency, the dust temperature $T_{\rm IR}$ can be computed by equating the dust heating by absorption with the dust cooling by emission. For the predominant UV radiation field, and using a dust emissivity law of the form $j = {\lambda}^{-2} B(T)$, where  $\lambda$ is the wavelength and $B(T)$ is Planck's emission law (Van Buren \& McCray 1988), we find: 
\begin{equation}
T_{\rm IR} = 27\,a_{\mu {\rm m}}^{-1/6}\,L_{{\star} 38}^{1/6}\,R_{0 {\rm pc}}^{-1/3} \,\, {\rm K}.
\label{Temp}
\end{equation}
Here $a_{\mu {\rm m}}$ $\sim$ $0.2$ $\mu$m is the dust grain radius, $R_{0 {\rm pc}}$ is $R_{0}$ in pc, and  $L_{{\star} 38}$ is the star luminosity in units $10^{38}$ erg s$^{-1}$. More complex dust emission models can be found in Draine \& Li (2007) and Draine (2011).

 A fraction  of the star's bolometric luminosity is re-emitted in the IR by the dust grains. The re-processed luminosity, $L_{\rm IR}$ , can be roughly  estimated as that of a black body at $T = T_{\rm IR}$. 

From Eqs. (\ref{R0}) and  (\ref{Temp}), it follows that $R_{0}$ $\propto$ $\rho_{\rm a}^{-1/2}$, and $T_{\rm IR}$ $\propto$ $\rho_{\rm a}^{1/6}$. Therefore, the shape, luminosity and temperature of the bowshock depend on the ambient density.

\section{Molecular cloud structure}\label{mc}

 Molecular clouds are the site of practically  all star formation in the Galaxy. Typical densities are $\sim$ $10^{2}-10^{3}$ cm$^{-3}$ (Crutcher et al. 2010). Here we adopt a value of $n_{\rm MC}$ $\sim$ $10^{2}$ cm$^{-3}$. These clouds have varied  structures on different length scales. The clouds collapse to form dense cores\footnote{A condensate structure of higher density than the average ($>3\sigma$) density in the MC.} through a combination of gravity and turbulence: gravoturbulent fragmentation (e.g., Klessen 2011). The higher-density structures  have typical temperatures of the order of $\sim$ 10 K and densities $\sim$ $10^4-10^5$ cm$^{-3}$ (Bodenheimer 2011).

\begin{figure}
\begin{center}
\resizebox{1.\columnwidth}{!}{\includegraphics[trim=0cm 0cm 0cm 0cm, clip=true,angle=270]{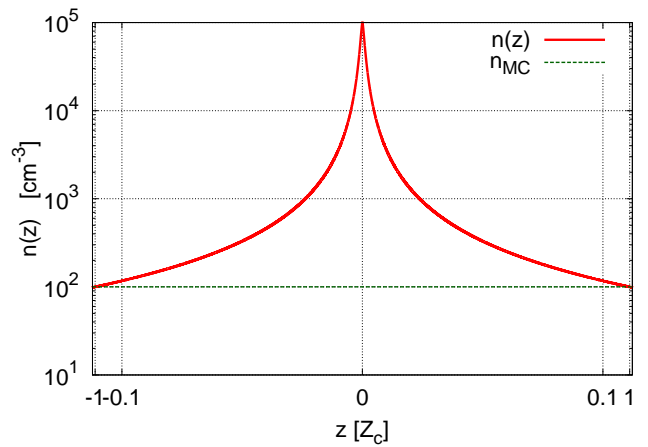}}
\caption{Density profile of the molecular cloud.}
\label{fig:profile}
\end{center}
\end{figure}

We consider a region in the MC with a plane-parallel density gradient of size $Z_{\rm c}$. The density profile is expected to be a power law (e.g., Smith, Clark \& Bonnelli 2009; Donkov, Veltchev \& Klessen 2011). We adopt a density profile of the form (see Fig. \ref{fig:profile}):
\begin{equation}
n(z) = \frac{n_{0}}{[1+ (z/Z_{\rm core})^{\delta}]}, 
\end{equation}
with $\delta$ $\sim$ $3/2$ (e.g., Smith, Clark \& Bonnelli 2009) and $Z_{\rm core}$ $\sim$ $10^{-2}Z_{\rm c}$. This value of $Z_{\rm core}$ ensures that $n(Z_{\rm c})$ $\rightarrow$ $n_{\rm MC}$. We adopt $n_{0}$ = $10^{5}\,\,{\rm cm}^{-3}$. 

\section{Radiative process in bowshock-medium interactions}\label{RP}

\begin{figure}
\begin{center}
\resizebox{1.\columnwidth}{!}{\includegraphics[trim=0cm 0cm 0cm 0cm, clip=true,angle=0]{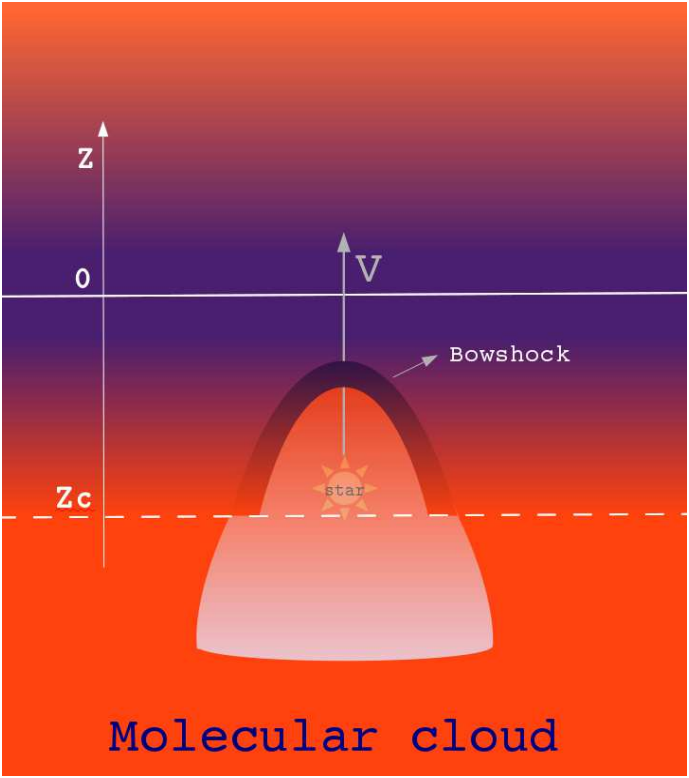}}
\caption{Simplified scheme of a runaway star moving through a denser region in a molecular cloud (not to scale).}
\label{fig:bow}
\end{center}
\end{figure}

We consider a bowshock of a runaway star that travels through a density gradient in an MC. As the star travels through the inhomogeneous  medium, the emission produced in the bowshock varies (see Fig. \ref{fig:bow} for a sketch of the situation).

\begin{table}
\begin{center}

\begin{tabular}{lll}
\hline\noalign{\smallskip}
 Parameter & O4I & O9I \\[0.001cm]
\hline\noalign{\smallskip}

$V_{\star}$  Spatial velocity [km s$^{-1}$] & 100 &  30\\[0.001cm]
$V_{\rm w}^{\rm a}$  Wind velocity [km s$^{-1}$] & 2.2$\times 10^3$ & 0.8$\times 10^3$   \\[0.001cm]
${\dot{M}}_{\rm w}^{\rm a}$  Wind mass loss rate [M$_{\odot}$ yr$^{-1}$] & $10^{-4}$ & $10^{-6}$ \\[0.001cm]
%$a$ Hadron-to-lepton energy ratio$^{\rm c}$ & 1 & 100\\[0.001cm]
$q_{\rm rel}$  Content of relativistic particles & 10\% & 10\% \\[0.001cm]
$\alpha$  Injection index & 2 & 2 \\[0.001cm]
${L_{\star}}^{\rm b}$ Star luminosity [$L_{\odot}$] & $\sim$ 7$\times 10^{5}$ & $\sim$ 5$\times 10^{4}$  \\[0.001cm]
${T_{\star}}^{\rm b}$ Star temperature [K] & 4.1$\times 10^{4}$ & 2.9$\times 10^{4}$  \\[0.001cm]
${R_{\star}}^{\rm b}$ Star radius [$R_{\odot}$] & $\sim$ 18.5 & $\sim$ 22.6  \\[0.001cm]

\hline\\[0.05cm]
\end{tabular}	
\caption[]{Parameters for the stars considered in this work, along with assumptions  related to relativistic particles accelerated at the reverse shock.\\
$^{\rm a}$Values derived by  Kobulnicky, Gilbert \& Kiminki (2010).\\
$^{\rm b}$Values from Martins, Schaerer \& Hillier (2005).\\
$^{\rm c}$At the radiation regions.
}
\label{tabla}
\end{center}
\end{table}

As mentioned before, the collision of the supersonic stellar wind with the ISM results in a system of two shocks (e.g., Wilkin 2000). Following the model developed by del Valle \& Romero (2012), we assume that relativistic particles are accelerated via the first-order Fermi mechanism in the reverse adiabatic shock. This shock propagates in the opposite  direction of the stellar motion,  with velocity $v_{\rm s}$ $\sim$ $V_{\rm wind}$. The stellar wind can be considered as a continuous power source, therefore both shocks, the forward and reverse shock, reach a steady state. We assume that the bowshock reaches a steady state almost immediately in its way through the density gradient, so a steady-state system can be considered for each value of $z$. 

We perform calculations for two types of massive stars: an O4I and an O9I star, as representative examples of a very powerful and a more modest case. Their adopted parameters are listed in Table \ref{tabla}. Many of the parameters that define the particle energy losses and the non-thermal emission change with $n$. In  Tables \ref{par-O4} and  \ref{par-O9}, the model parameters as a function of $z$ (i.e., $n$) are listed for both bowshocks. The shape of the bowshock surface also changes with $n$, but our model is not sensitive to these changes. We consider a one-zone homogeneous cap region where particles are accelerated  and emit radiation. This region is located near the apex of the bowshock, where the shock is nearly planar. 

\begin{table*}
\begin{center}
\begin{tabular}{llllll}
%\begin{tabular}{|c|c|c|c|c|c|c|}{lllll}
\hline\noalign{\smallskip}
%\hline
%\multicolumn{2}{|c|}{Ene} & {Ene} \\
$z$ & $n$ & $R_{0}$ & $B$ & $T_{\rm IR}$ & $E_{\rm e\,max}$\\[0.001cm]
\hline\noalign{\smallskip}
$Z_{\rm c}$ & $10^{2}$ cm$^{-3}$ & $5.4\times 10^{-1}$ pc & $4.49\times 10^{-4}$ G & 60 K & $2.5\times10^{12}$ eV\\[0.001cm]
%\hline
$10^{-1}Z_{\rm c}$ & $3\times10^{3}$ cm$^{-3}$ & $9.8\times 10^{-2}$ pc & $2.46\times 10^{-3}$ G & 106 K & $1.0\times10^{12}$ eV\\[0.001cm]
%\hline
$10^{-2}Z_{\rm c}$ & $5\times10^{4}$ cm$^{-3}$ & $2.4\times 10^{-2}$ pc & $1.\times 10^{-2}$ G & 170 K & $5.4\times10^{11}$ eV\\[0.001cm]
%\hline
$\sim 0$ & $10^{5}$ cm$^{-3}$ & $1.7\times 10^{-2}$ pc & $1.42\times 10^{-2}$ G & 190 K & $4.5\times10^{11}$ eV\\[0.001cm]
\hline\\[0.05cm]
\end{tabular}
\end{center}
\caption{Model parameters as a function of $z$ for an O4I star.}
\label{par-O4}
\end{table*}

\begin{table*}
\begin{center}
\begin{tabular}{llllll}
%\begin{tabular}{|c|c|c|c|c|c|c|}{lllll}
\hline\noalign{\smallskip}
%\begin{tabular}{|c|c|c|c|c|c|c|}
%\hline
%\multicolumn{2}{|c|}{Ene} & {Ene} \\
$z$ & $n$ & $R_{0}$ & $B$ & $T_{\rm IR}$ & $E_{\rm e\,max}$ \\[0.001cm]
\hline
$Z_{\rm c}$ & $10^{2}$ cm$^{-3}$ & $10^{-1}$ pc & $1.35\times 10^{-4}$ G & 66 K & $5.0\times10^{11}$ eV\\[0.001cm]
%\hline
$10^{-1}Z_{\rm c}$ & $3\times10^{3}$ cm$^{-3}$ & $1.97\times 10^{-2}$ pc & $7.38\times 10^{-4}$ G & 116 K & $3.2\times10^{11}$ eV\\[0.001cm]
%\hline
$10^{-2}Z_{\rm c}$ & $5\times10^{4}$ cm$^{-3}$ & $4.8\times 10^{-3}$ pc & $3.01\times 10^{-3}$ G & 186 K & $2.8\times10^{11}$ eV\\[0.001cm]
%\hline
$\sim 0$ & $10^{5}$ cm$^{-3}$ & $3.41\times 10^{-3}$ pc & $4.26\times 10^{-3}$ G & 209 K & $2.5\times10^{11}$ eV\\[0.001cm]
\hline\\[0.05cm]
\end{tabular}
\end{center}
\caption{Model parameters as a function of $z$ for an O9I star.}
\label{par-O9}
\end{table*}
 
The  acceleration timescale as a function of the  energy $E$,  for a  charged particle being accelerated in a magnetic field $B$, is given by (e.g., Gaisser 1990; Aharonian 2004; Bosch-Ramon 2009; Romero \& Paredes 2011):
\begin{equation}
t_{\rm acc} = {\eta} \frac{r_{\rm L}}{c} = \eta \frac{E}{eBc},
\label{acelera}
\end{equation}   
where $r_{\rm L}$ is the Larmor radius $r_{\rm L} = E/eB$, and $\eta$  is a phenomenological parameter related to the efficiency of the acceleration process involved. For a non-relativistic diffusive shock acceleration, in a plane shock in the test particle approximation $\eta$ can be approximated by (Drury 1983):

\begin{equation}
\eta \sim {20}\, \frac{D}{r_{\rm g}\,c}\left(\frac{c}{v_{\rm s}} \right)^{2}, 
\end{equation}
where $D$ is the diffusion coefficient and $r_{\rm g} = E/(eB)$ is the particle gyro-radius. In the Bohm limit, $D = D_{\rm B}$ and $D_{\rm B} = r_{\rm g}c/3$  so $\eta$ becomes  $\eta \sim {20}/{3}\left({c}/{v_{\rm s}} \right)^{2}$.
 
We estimate the magnetic field  considering  that the magnetic energy density is in  sub-equipartition with respect to the kinetic energy $L_{\rm T} \sim  \frac{1}{2}{\dot M_{\rm w}}V_{\rm w}^2$,  by a 0.1 factor, i.e.,
\begin{equation}
\frac{B_{\rm sub}^{2}}{8\pi} = \frac{0.1 L_{\rm T}}{V_{\rm w} A},
\label{equipar}
\end{equation}
where $A$ is the area of a sphere of radius $R_{0}$. This guarantees that the plasma remains compressible and shocks are not suppressed by the magnetic fields.  The magnetic field in the acceleration region is $B$ $\sim$ $4B_{\rm sub}$ due to the compression by the shock.

\subsection{Energy losses}

\begin{figure*}
\begin{center}
\resizebox{.8\hsize}{!}{\includegraphics[trim= 1.5cm 2cm 1.5cm 1cm, clip=true,angle=270]{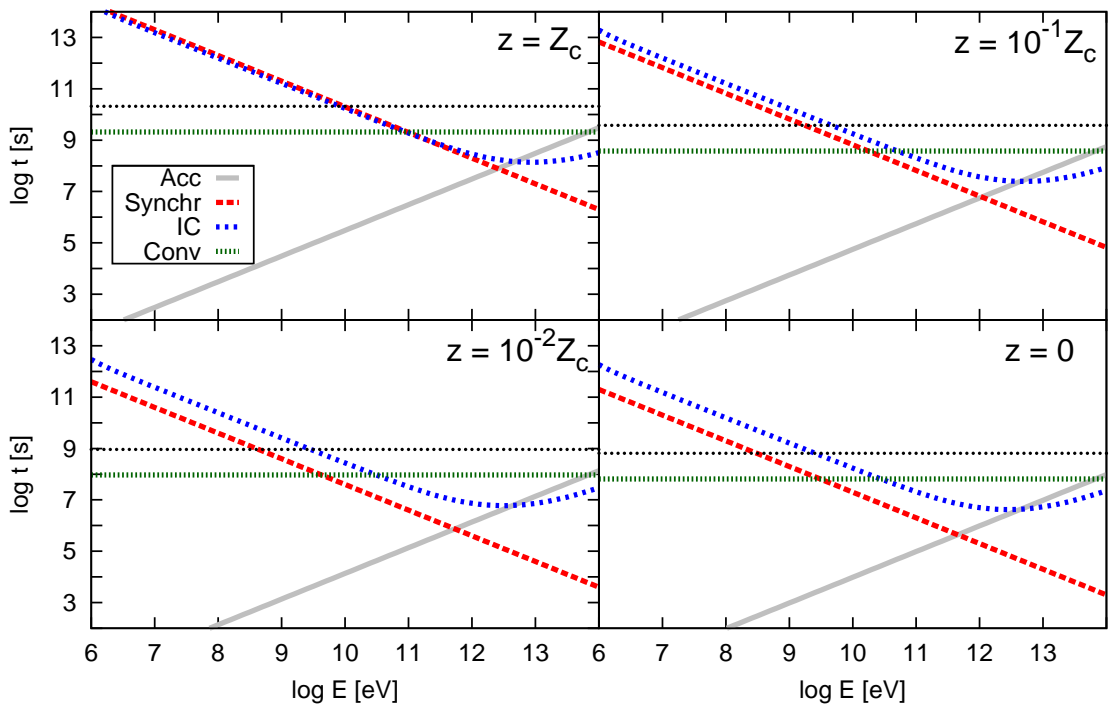}}
\caption{Electron losses for $z$ = $0$, $10^{-2}Z_{\rm c}$, $10^{-1}Z_{\rm c}$, and $Z_{\rm c}$, for a O4I star. The dotted line corresponds to slow convection time.}
\label{fig:cool1}
\end{center}
\end{figure*}

\begin{figure*}
\begin{center}
\resizebox{.8\hsize}{!}{\includegraphics[trim=1.5cm 2cm 1.5cm 1cm, clip=true,angle=270]{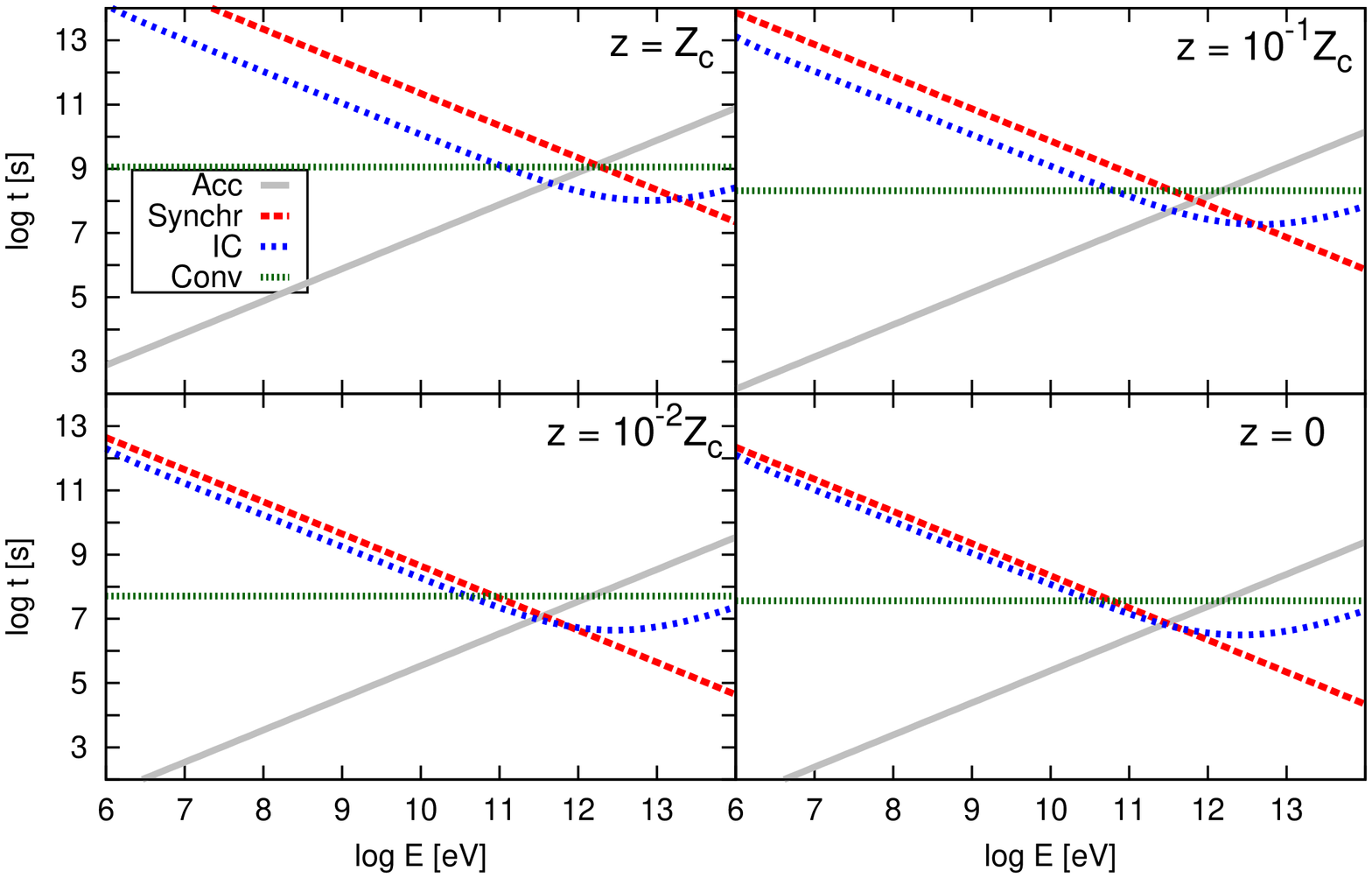}}
\caption{Electron losses for $z$ = $0$, $10^{-2}Z_{\rm c}$, $10^{-1}Z_{\rm c}$, and $Z_{\rm c}$, for a O9I star.}
\label{fig:cool2}
\end{center}
\end{figure*}

We calculate the energy losses at different radii $z$ = $0$, $10^{-2}Z_{\rm c}$, $10^{-1}Z_{\rm c}$, and $Z_{\rm c}$, as the star moves through the density gradient. It is clear that this scenario is symmetric: after reaching the maximum density at $z$ = $0$, the density decreases to $n_{\rm MC}$. Particles lose energy when interacting with the magnetic fields, radiation, and matter. The electrons lose energy mainly by IC scattering, synchrotron radiation, and relativistic Bremsstrahlung. Protons cool through proton-proton in-elastic collisions with the  ambient gas, but they escape from the radiation region convected away by the stellar wind. The convection time might be longer than $t_{\rm conv} \sim \Delta/V_{\rm W}$ (here $\Delta$ is the width of the shocked wind) due to turbulence in the flow driven by instabilities. Therefore, for the O4I star, we consider a regular convection time, \emph{case a}, and we also consider a longer convection time, \emph{case b}, which is one order of magnitude longer than $t_{\rm conv}$. For the mathematical expressions of the losses, see del Valle \& Romero 2012 and references therein.

Since convection imposes the upper limit to the energy of protons for all $z$, the maximum energy that these particles can reach almost remains unchanged with $z$. For the O4I star, the highest energies protons reach are                                   $\sim$ 70 TeV and 700 TeV for \emph{case a} and \emph{case b}, respectively. For the O9I star, protons are accelerated up to $\sim$ 1 TeV.
 Most of the protons escape without losing much of their energy, and they might produce non-thermal radiation further away in the cloud (del Valle \& Romero, in preparation). The dominant loss for electrons changes with  $z$, therefore, the maximum  energy that electrons can achieve varies for each $z$, in general, decreasing as $z$ decreases. In Fig. \ref{fig:cool1} and \ref{fig:cool2}, we show the most relevant energy losses, at $z$ = $0$, $10^{-2}Z_{\rm c}$, $10^{-1}Z_{\rm c}$, and $Z_{\rm c}$, for the O4I and O9I stars, respectively. 

\subsection{Particle energy distributions}

We compute the particle energy distribution for both species of particles, solving the  steady state transport equation in the homogeneous approximation  (Ginzburg \& Syrovatskii 1964):

 \begin{equation}
 \frac{\partial}{\partial E}\biggl[\frac{{\rm d}E}{{\rm d}t}{\bigg\arrowvert}_{\rm loss}N(E)\biggr]+\frac{N(E)}{t_{\rm esc}} = Q(E).
\label{tra}
\end{equation}
Here $Q(E)$ is the power-law injection function, $t_{\rm esc}$ is the wind convection time, and $({\rm d} E/{\rm d}t)_{\rm loss}$ are the radiative losses (see del Valle \& Romero 2012, for further details). 

\begin{figure}[h!]
\begin{center}
\resizebox{1.\columnwidth}{!}{\includegraphics[trim=0cm 0cm 1.5cm 0cm, clip=true,angle=270]{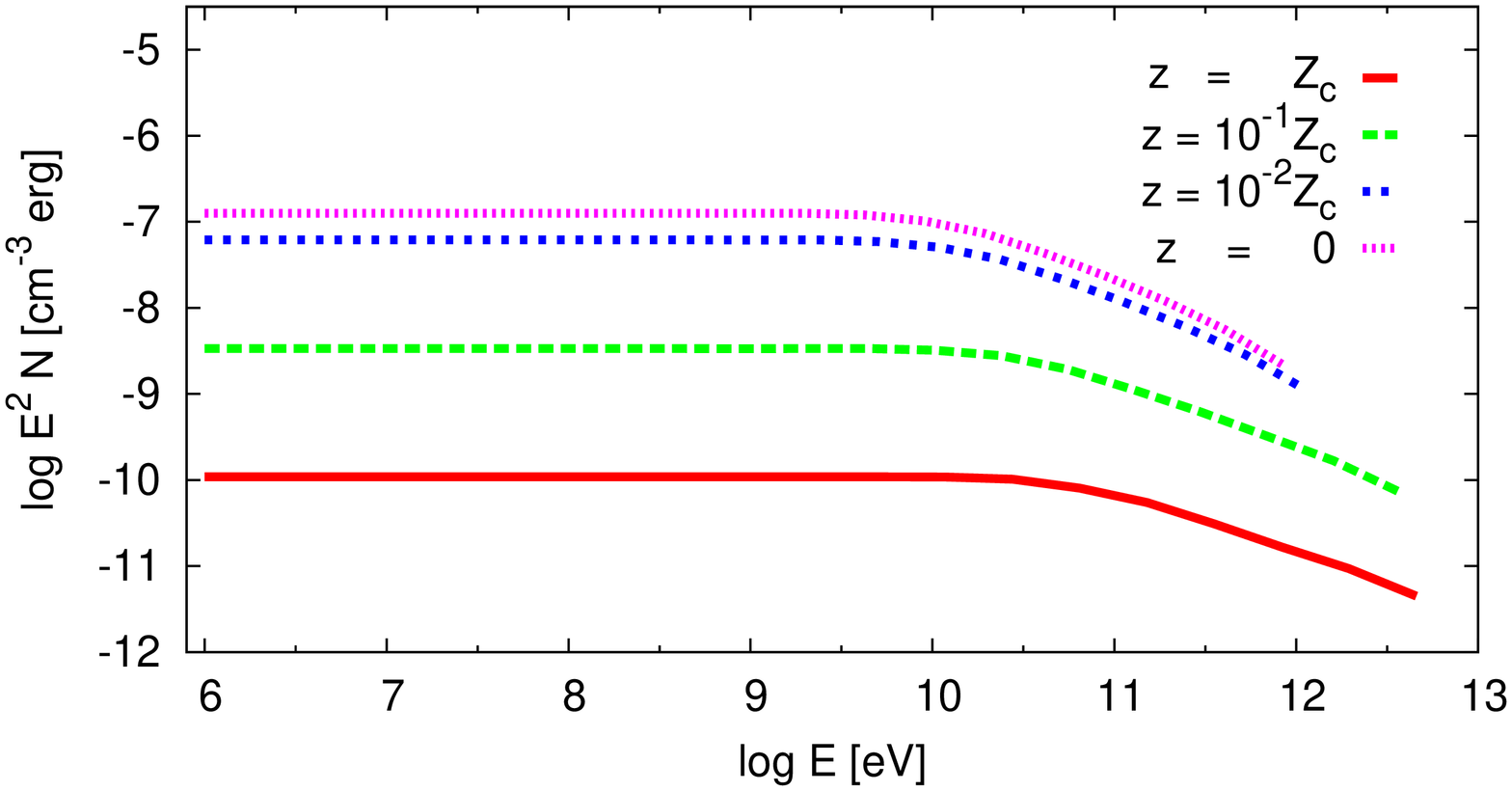}}
\resizebox{1.\columnwidth}{!}{\includegraphics[trim=1.5cm 0cm 0cm 0cm, clip=true,angle=270]{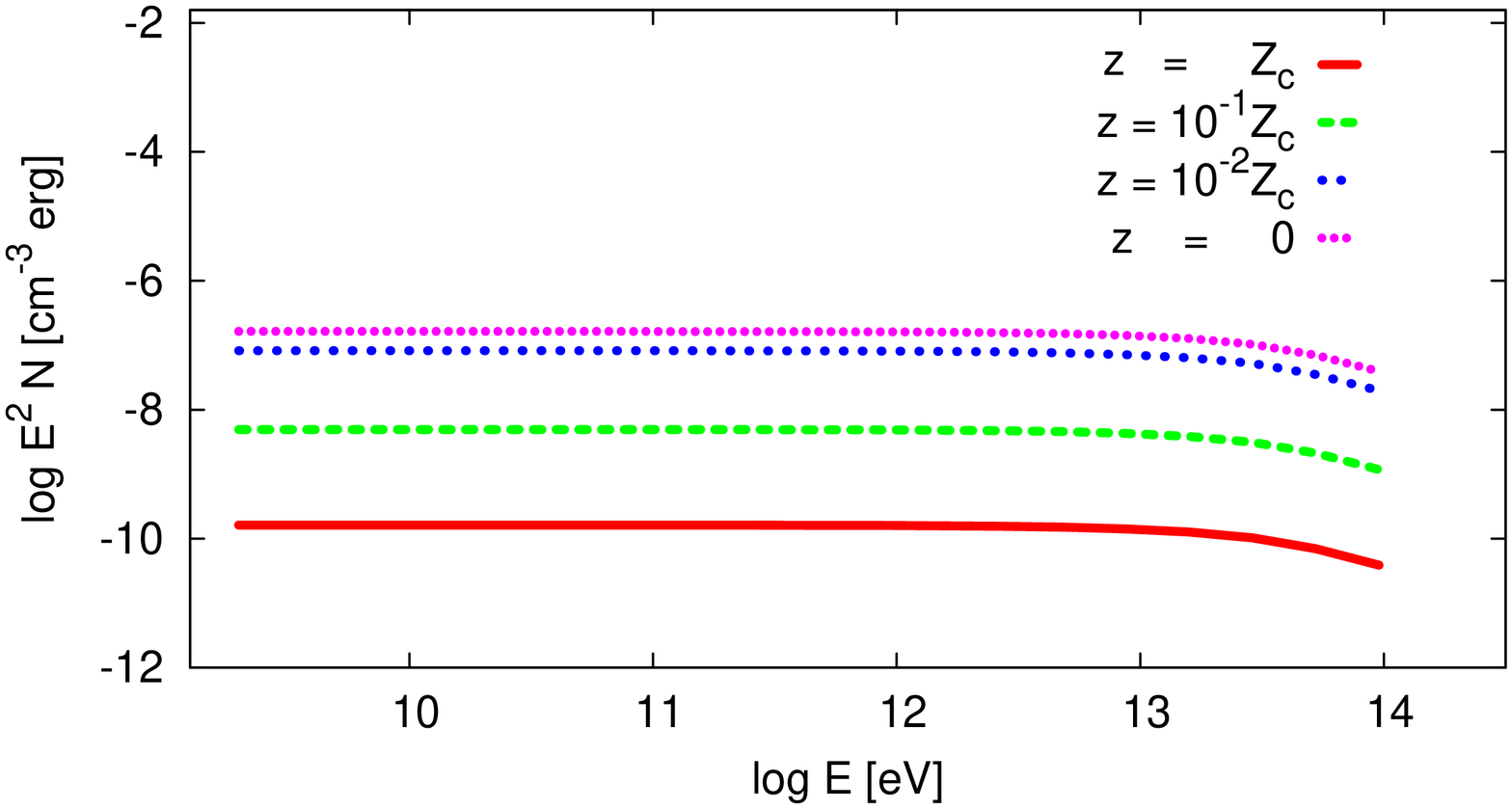}}
\caption{Particle distribution  for electrons (up) and protons (down), for the  O4I star in \emph{case a}, at  $z$ = $0$, $10^{-2}Z_{\rm c}$, $10^{-1}Z_{\rm c}$, and $Z_{\rm c}$.}
\label{N-O4}
\end{center}
\end{figure}

\begin{figure}[h!]
\begin{center}
\resizebox{1.\columnwidth}{!}{\includegraphics[trim=0cm 0cm 1.5cm 0cm, clip=true,angle=270]{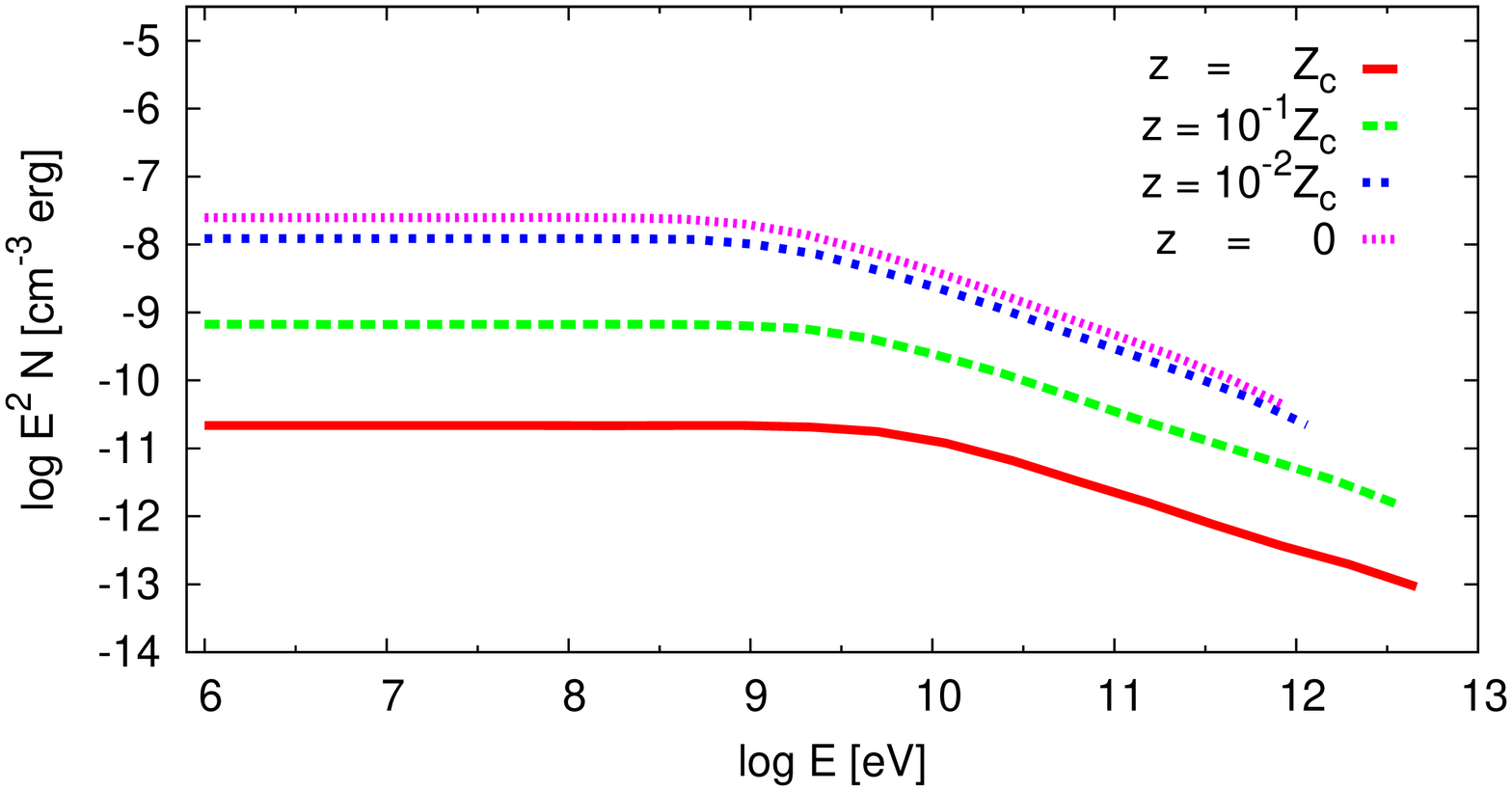}}
\resizebox{1.\columnwidth}{!}{\includegraphics[trim=1.5cm 0cm 0cm 0cm, clip=true,angle=270]{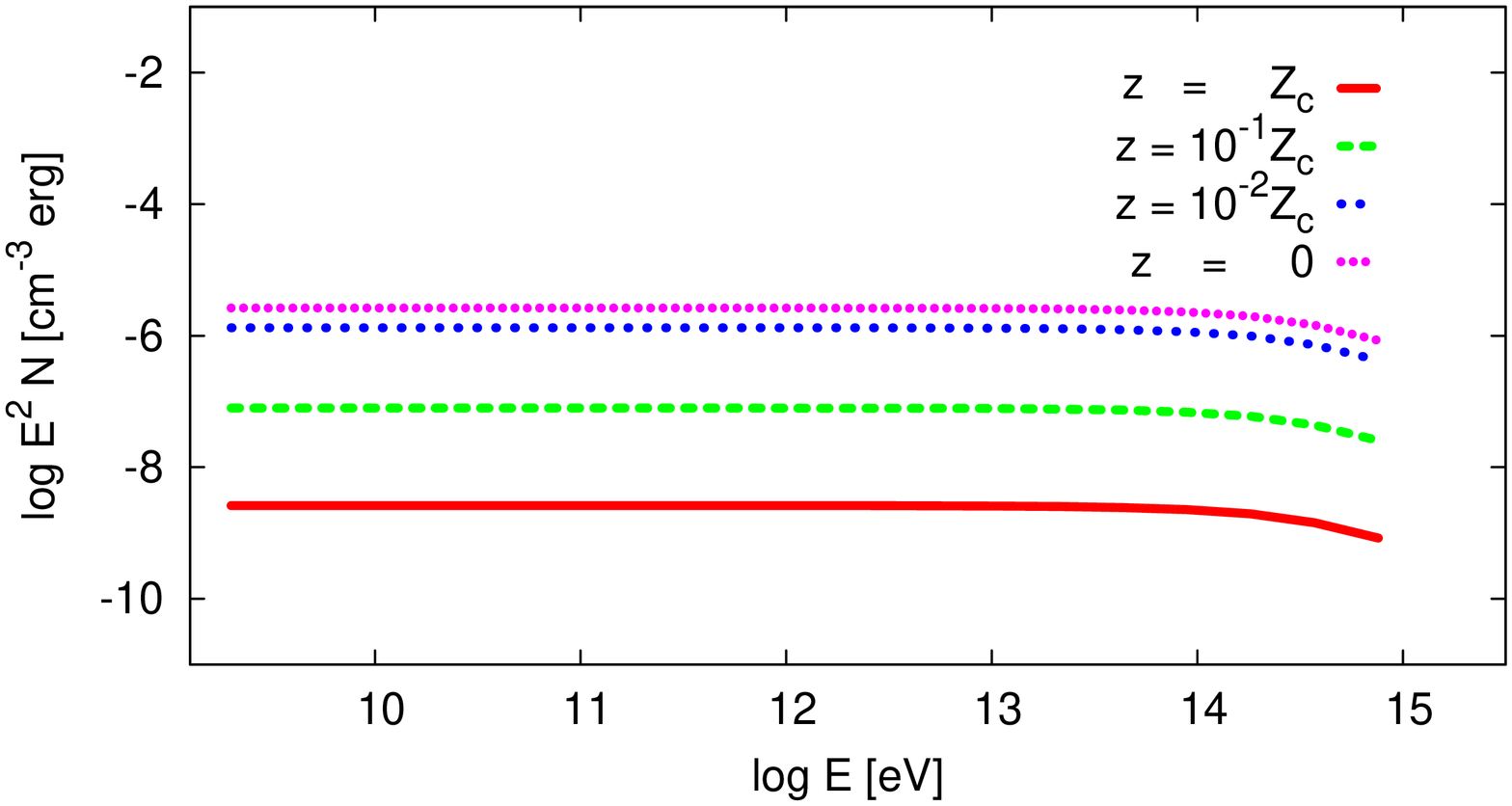}}
\caption{Particle distribution  for electrons (up) and protons (down), for the  O4I star in \emph{case b}, at  $z$ = $0$, $10^{-2}Z_{\rm c}$, $10^{-1}Z_{\rm c}$, and $Z_{\rm c}$.}
\label{N-O4b}
\end{center}
\end{figure}

\begin{figure}[h!]
\begin{center}
\resizebox{1.\columnwidth}{!}{\includegraphics[trim=0cm 0cm 1.5cm 0cm, clip=true,angle=270]{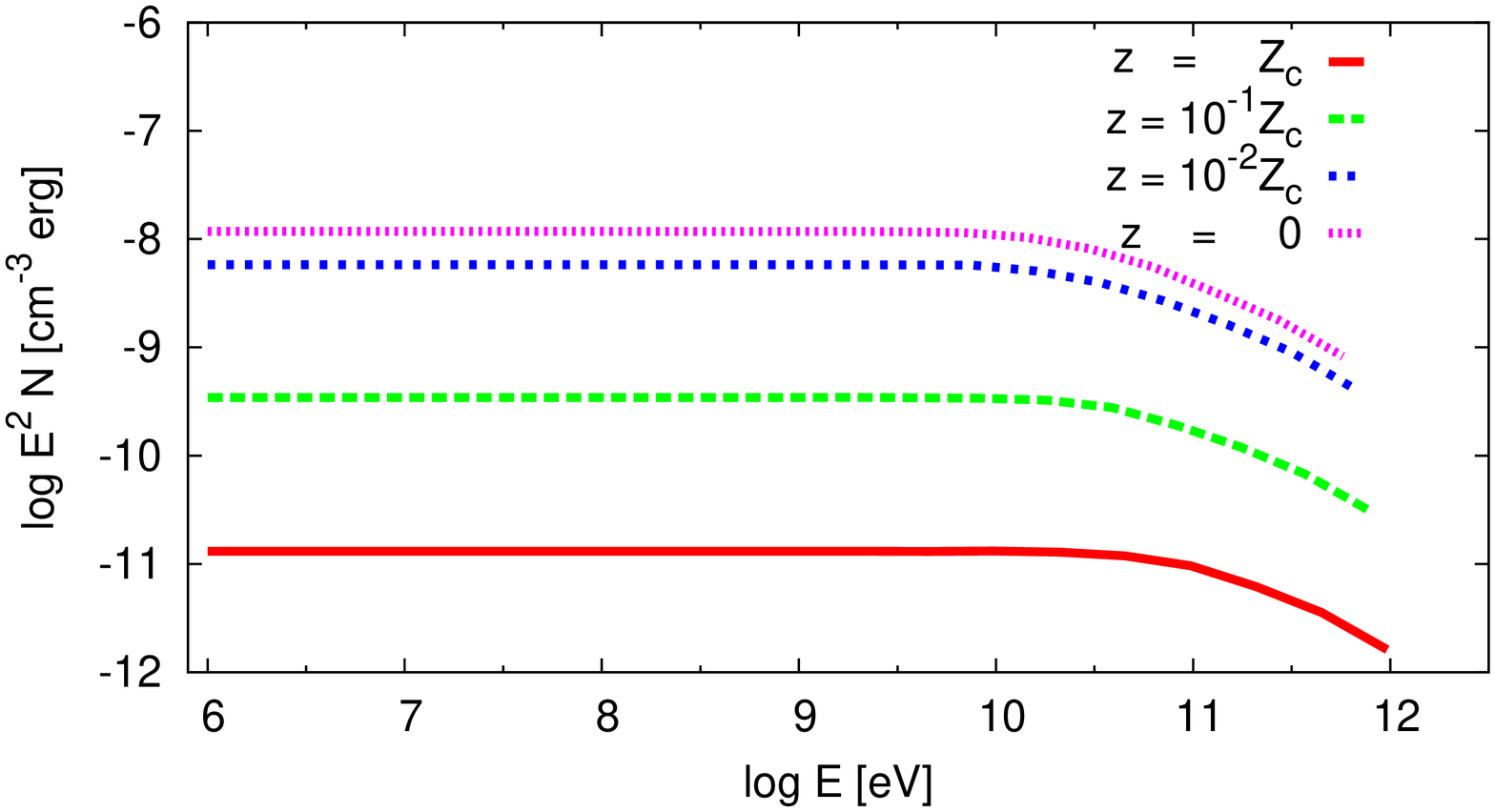}}
\resizebox{1.\columnwidth}{!}{\includegraphics[trim=1.5cm 0cm 0cm 0cm, clip=true,angle=270]{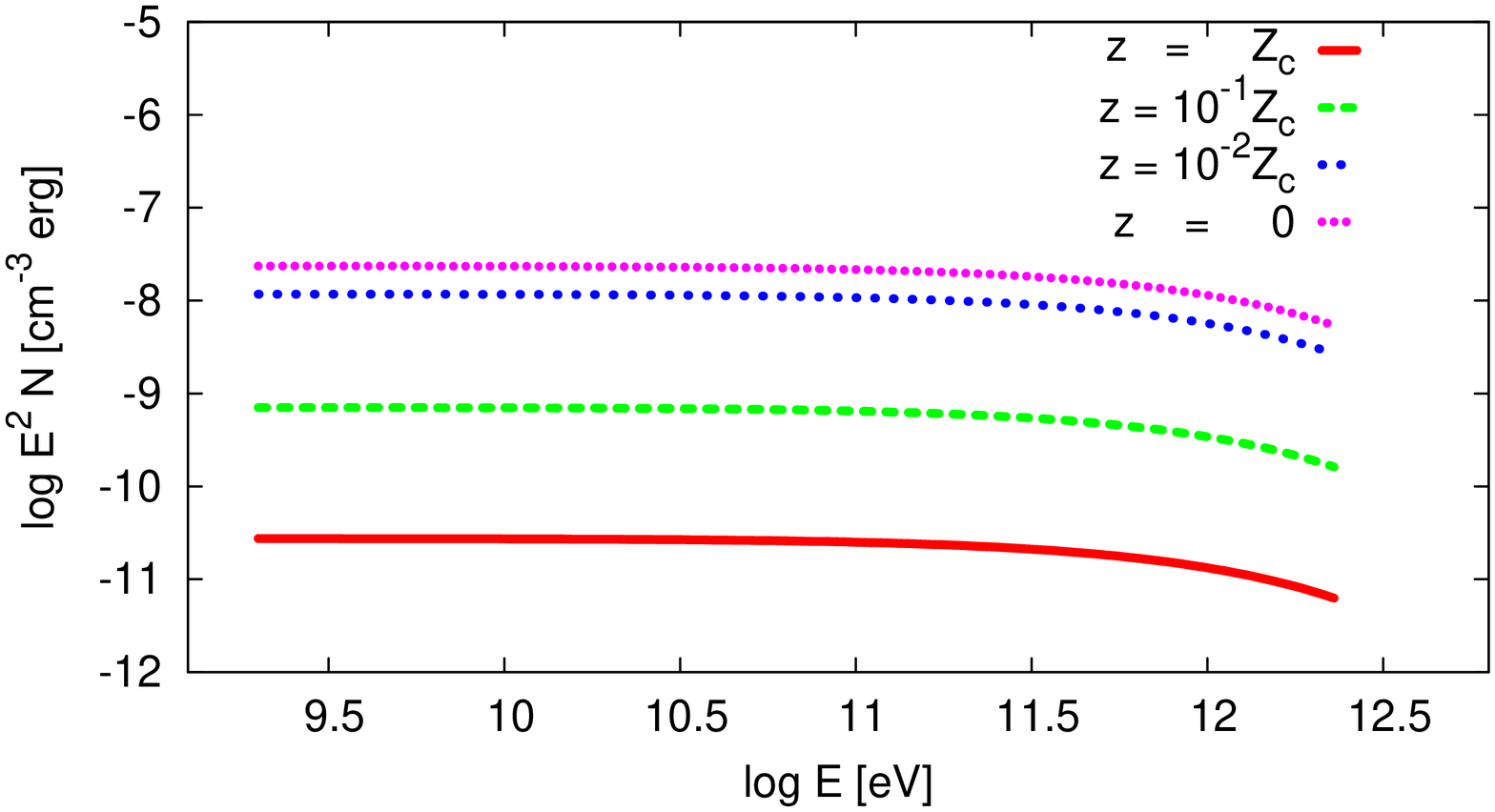}}
\caption{Particle distribution for electrons (up) and protons (down), for the  O9I star, at  $z$ = $0$, $10^{-2}Z_{\rm c}$, $10^{-1}Z_{\rm c}$, and $Z_{\rm c}$.}
\label{N-O9}
\end{center}
\end{figure}

The solution of Eq~(\ref{tra}) is a broken power law in the particle energy.
The ratio of relativistic proton power to relativistic  electron power, $a$, is unknown. We consider $a = 1$ (equal energy density in both particle species) for the O4I star in \emph{case a}, and for the O9I star; and $a = 100$ (as observed in galactic cosmic rays, Ginzburg \& Syrovatskii 1964), for the O4I star in \emph{case b}.
Figures  \ref{N-O4}, \ref{N-O4b}, and \ref{N-O9} show the computed particle distributions as a function of energy per unit volume for both electrons and protons, respectively. The normalisation constant of the injection function changes with $z$, and because of this, the particle  distributions change.


\subsection{Spectral energy distributions}

We compute the non-thermal luminosity  for different values of $z$ as the star moves through the density gradient. Figures \ref{all-O4} and \ref{all-O9} show the SEDs at the different locations of the stars. In the case of star O4I in \emph{case a}, the synchrotron and IC of IR photons  are strong and dominate the SEDs for energies $E < $ 1 TeV. The IC cut-off  decreases with $z$, and as $n$ increases, the $p-p$ component gets stronger, dominating the SEDs for $1<E<10^{2}$ TeV. For \emph{case b}, all leptonic contributions are weak; for $z = Z_{\rm c}$, the synchrotron and IC of IR photons dominate the SED up to the IC cut-off; as the star moves further in the MC, the hadronic contribution becomes stronger at high energies, and dominates the SEDs for $E>$ MeV.  In the case of the O9I system,  the SEDs are dominated by leptonic contributions at high energies, while the synchrotron emission is weak. In the range $z = 10^{-2}-0$, the IC of stellar photons dominates the SEDs for $E>1$ MeV because as $n$ grows $R_{0}$ decreases and the emission region gets closer to the strong stellar photon field. In all cases, the synchrotron's lower cut-off shifts to higher energies due to  the synchrotron's self absorption; this effect becomes important as the emission region becomes more compact.

\subsection{Absorption}

\begin{figure*}[htb!]
\begin{center}
\resizebox{.6\textwidth}{!}{\includegraphics[trim=.8cm 0.cm .8cm 0cm, clip=true,angle=270]{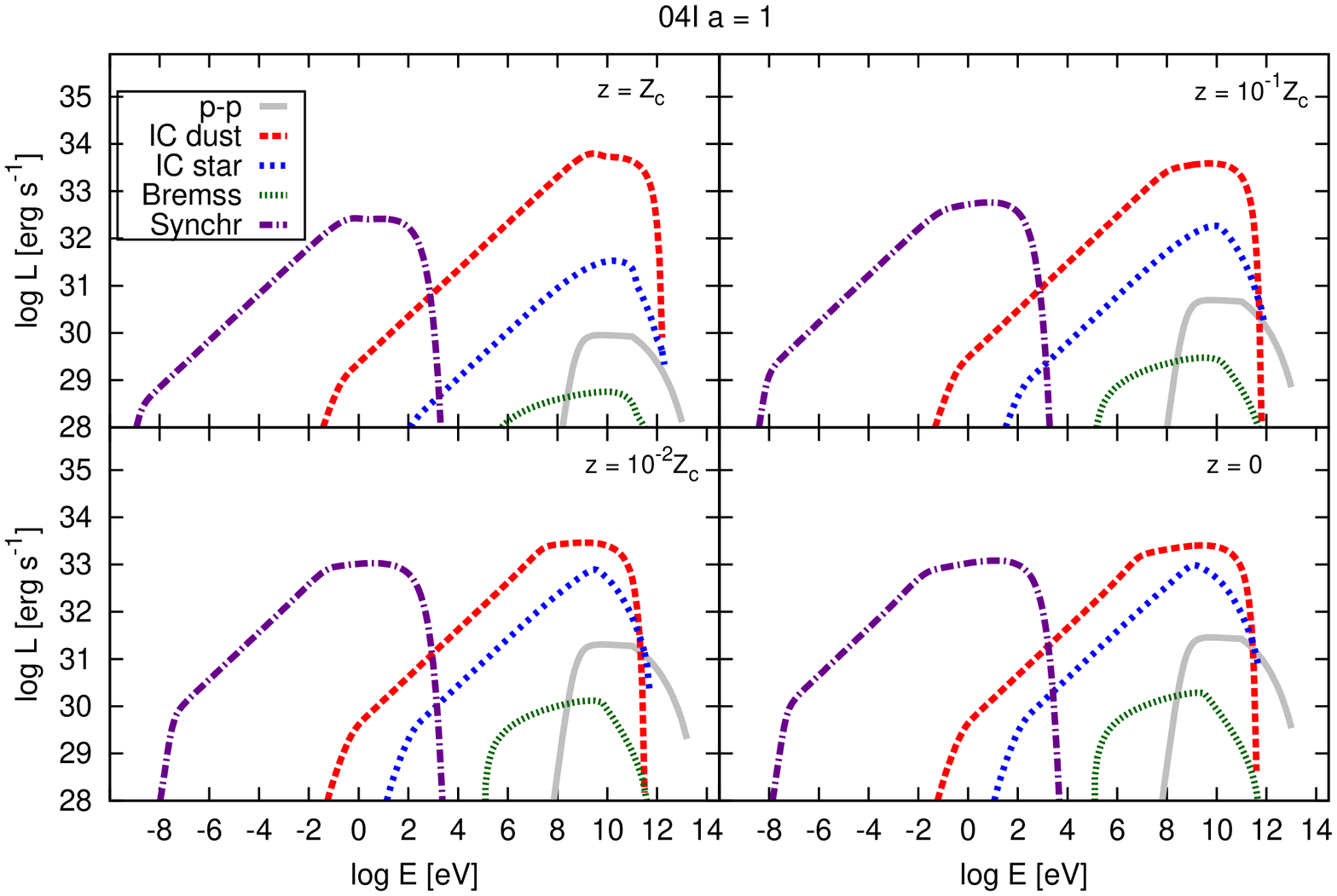}}
\resizebox{.6\textwidth}{!}{\includegraphics[trim=.8cm 0.cm 0cm 0cm, clip=true,angle=270]{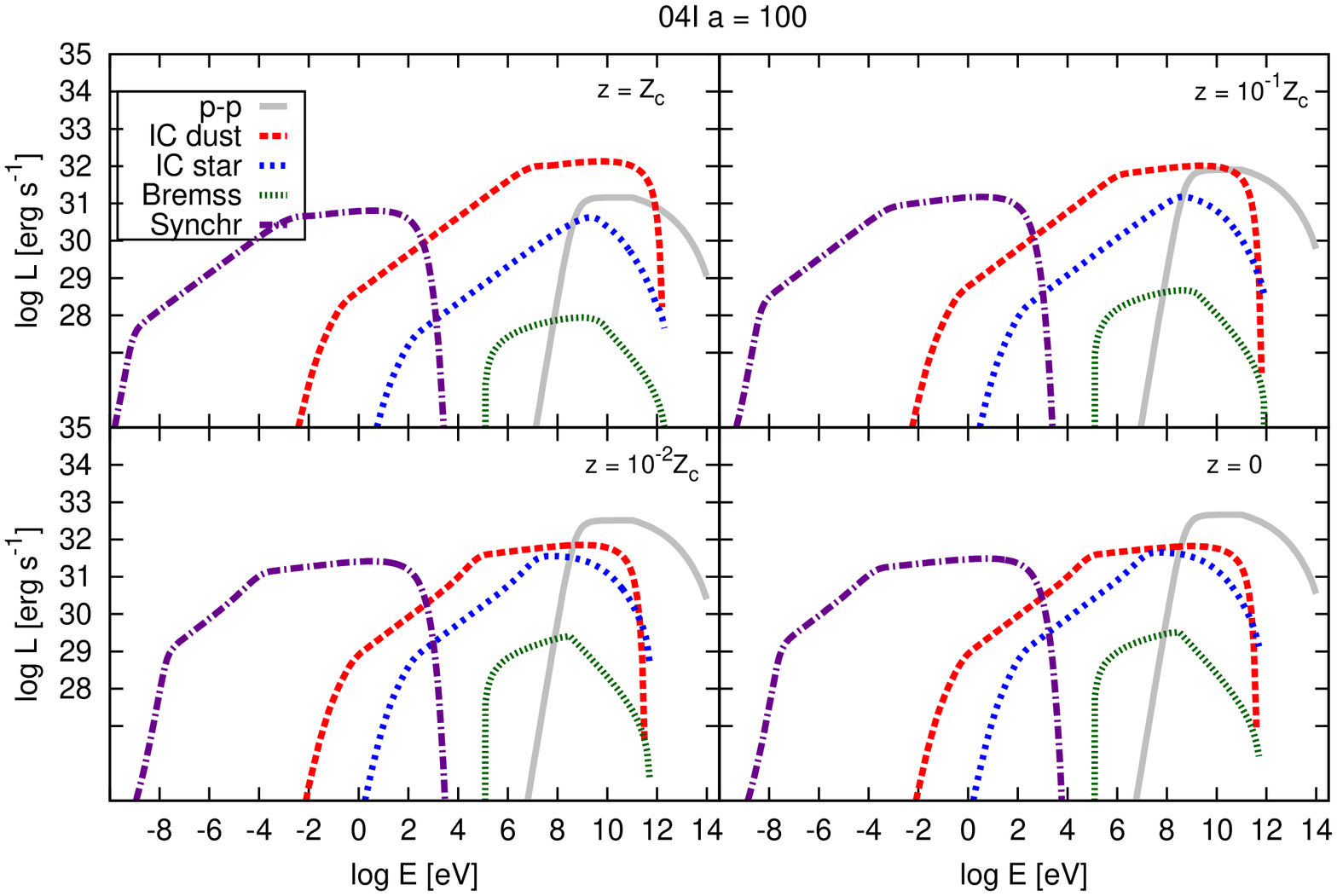}}
\caption{The SEDs for an O41-type star at $z$ = $0$, $10^{-2}Z_{\rm c}$, $10^{-1}Z_{\rm c}$, and $Z_{\rm c}$, \emph{case a} (top) and \emph{case b} (bottom).}
\label{sed-O4}
\end{center}
\end{figure*}

\begin{figure*}[htb!]
\begin{center}
\resizebox{.6\textwidth}{!}{\includegraphics[trim=0cm .8cm 0cm 0cm, clip=true,angle=270]{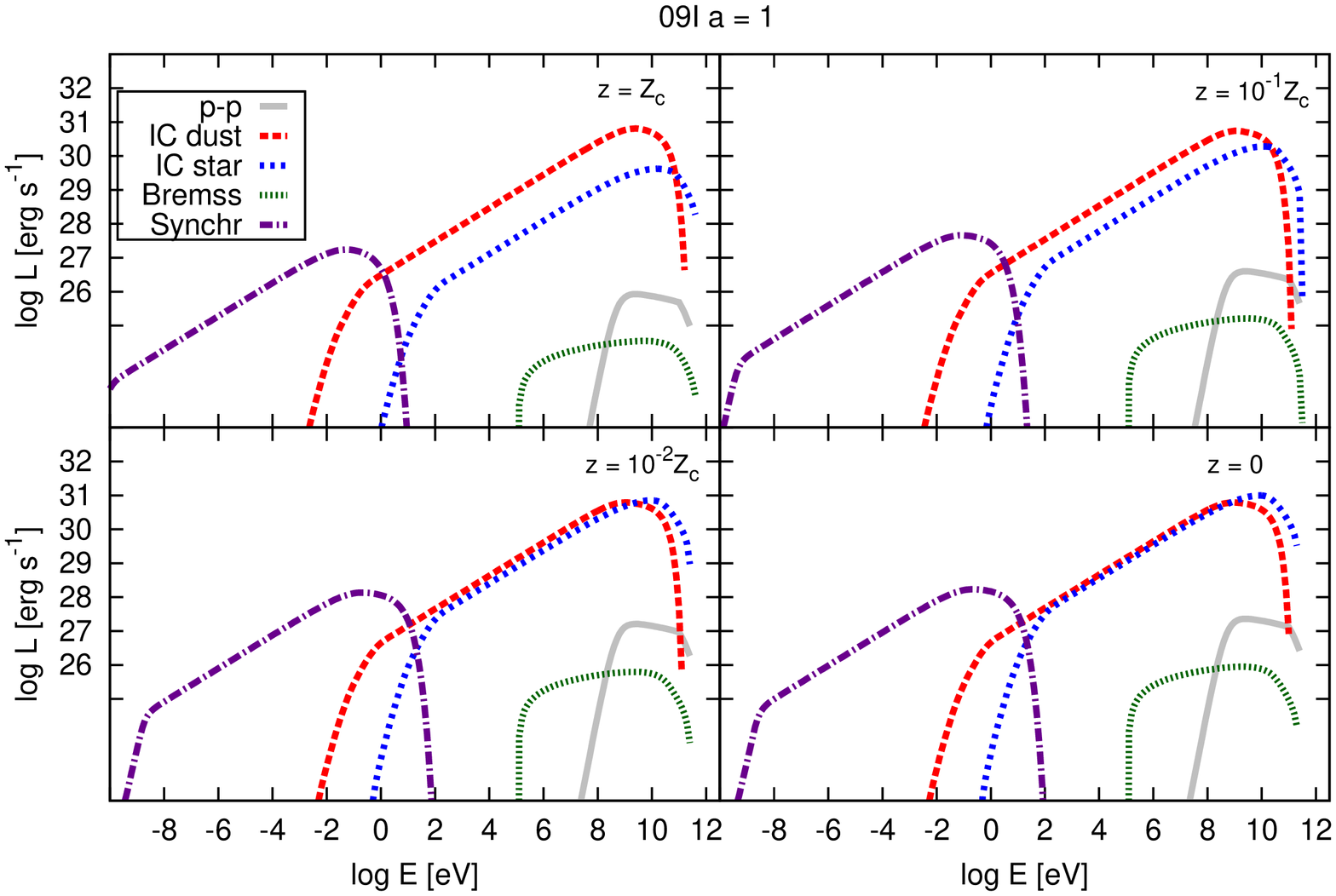}}
%\resizebox{1.\columnwidth}{!}{\includegraphics[trim=0cm 0cm 0cm 0cm, clip=true,angle=270]{multiplot-sed-O9-a100.eps}}
\caption{The SEDs for an O91-type star at $z$ = $0$, $10^{-2}Z_{\rm c}$, $10^{-1}Z_{\rm c}$, and $Z_{\rm c}$. }
\label{sed-O9}
\end{center}
\end{figure*}

Photons can be absorbed by different mechanisms inside the emitting region, called {\it internal absorption} and in their way to the observer, {\it external absorption}. The internal absorption is produced via photon-photon pair production; the absorbing photon fields are: the IR, non-thermal, and  stellar fields (see del Valle \& Romero 2012). The external absorption is produced by the matter fields and the star photon field. This latter contribution depends on the inclination  angle $i$ with the line of sight; the closest distance to the star is given by $R_{0}\sin{i}$, i.e., the absorption produced by this component would be non-negligible only in the particular case when $i$ $\simeq$ 0 (e.g., Romero, del Valle \& Orellana 2010). 

For energies $<$ 10 keV, the absorption in the ambient material is important and catastrophic for a wide range of energies. The photons must travel through the dense MC before reaching the observer. They can be absorbed  by photo-ionization for $E_{\gamma}$ $<$ 13.6 eV, and scattered by dust for lower energies, in the IR up to the ultra-violet (UV), (e.g., Ryter 1996; Reynoso, Medina \& Romero 2011). We estimate this opacity as:
\begin{equation}
\tau_{\gamma \rm N} \simeq N_{\rm H} \sigma_{\gamma \rm N}
\end{equation} 
where $\sigma_{\gamma \rm N}$ is the interaction cross section (see, e.g., Reynoso, Medina \& Romero 2011) and $N_{\rm H}$ is the column density of the MC. We adopt here a typical value $N_{\rm H}$ $=$ $1.5\times 10^{22}$ cm$^{-2}$ (e.g., Solomon et al. 1987). 
The denser region in the MC does not produce significant absorption in $\gamma$-rays because its maximum column density is
$\sim$ $n_{\rm c} Z_{\rm c}$ $<$ $10^{22}$ cm$^{-2}$, for $z_{c}$ $<$ $0.1$ pc. The same situation happens with the shocked ISM: it can be very dense, but it is confined to a very thin region.

\subsection{Results}

In Figures \ref{all-O4} and \ref{all-O9}, the non-thermal luminosity vs energy curves corrected by absorption  are shown. In these figures, we also indicate the 1-yr {\it Fermi} sensitivity curves for different distances. The non-thermal emission increases with $n$  at radio and X ray energies for both stars. The gamma-ray emission also increases, except for the star O4 in \emph{case a}.  For the O4 star, both the radio and the gamma-ray contributions are comparable. In the case of the O9 star, the high-energy emission dominates the spectral energy distribution (SED). This is because the magnetic field $B$ is relatively low, and the IR photon field is strong (this field is the main target of the IC scattering). The emission in the energy range between the near IR to the soft X-rays is fully suppressed. Internal absorption is only non-negligible at the tail of the high-energy spectrum. 

\section{Discussion: variability}\label{end}

In the scenario presented in this paper, the bowshock itself (manifested mainly by the IR signal) might not be detected because of the obscuring MC. Soft non-thermal X-rays  might also be difficult to detect due to high absorption and contamination by stronger thermal radiation. However, the radio and gamma-ray emission are not affected by the MC absorption and might be detectable. Radio emission is strong for the O4I star in \emph{case a}; and gamma-ray emission dominates the energy output in both types of stars.     
The variation on the gamma emission at TeV energies for the O4 star, in \emph{case a} and  \emph{case b}, is significant due to the gradual increase of the hadronic contribution to the SED.

\begin{figure}
\begin{center}
\resizebox{1.\columnwidth}{!}{\includegraphics[trim=0cm .8cm 0cm 0cm, clip=true,angle=270]{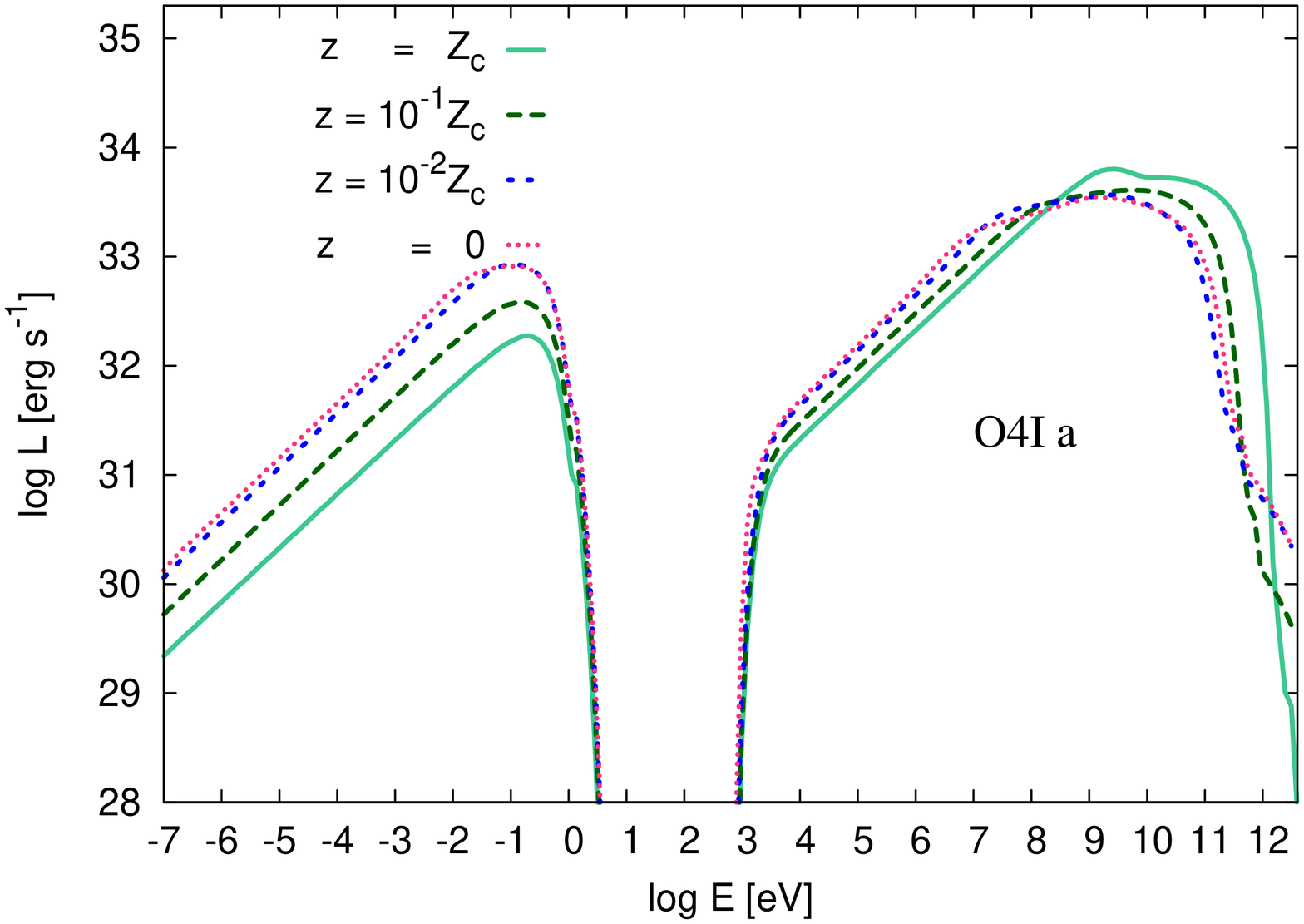}}
\resizebox{1.\columnwidth}{!}{\includegraphics[trim=0cm .8cm 0cm 0cm, clip=true,angle=270]{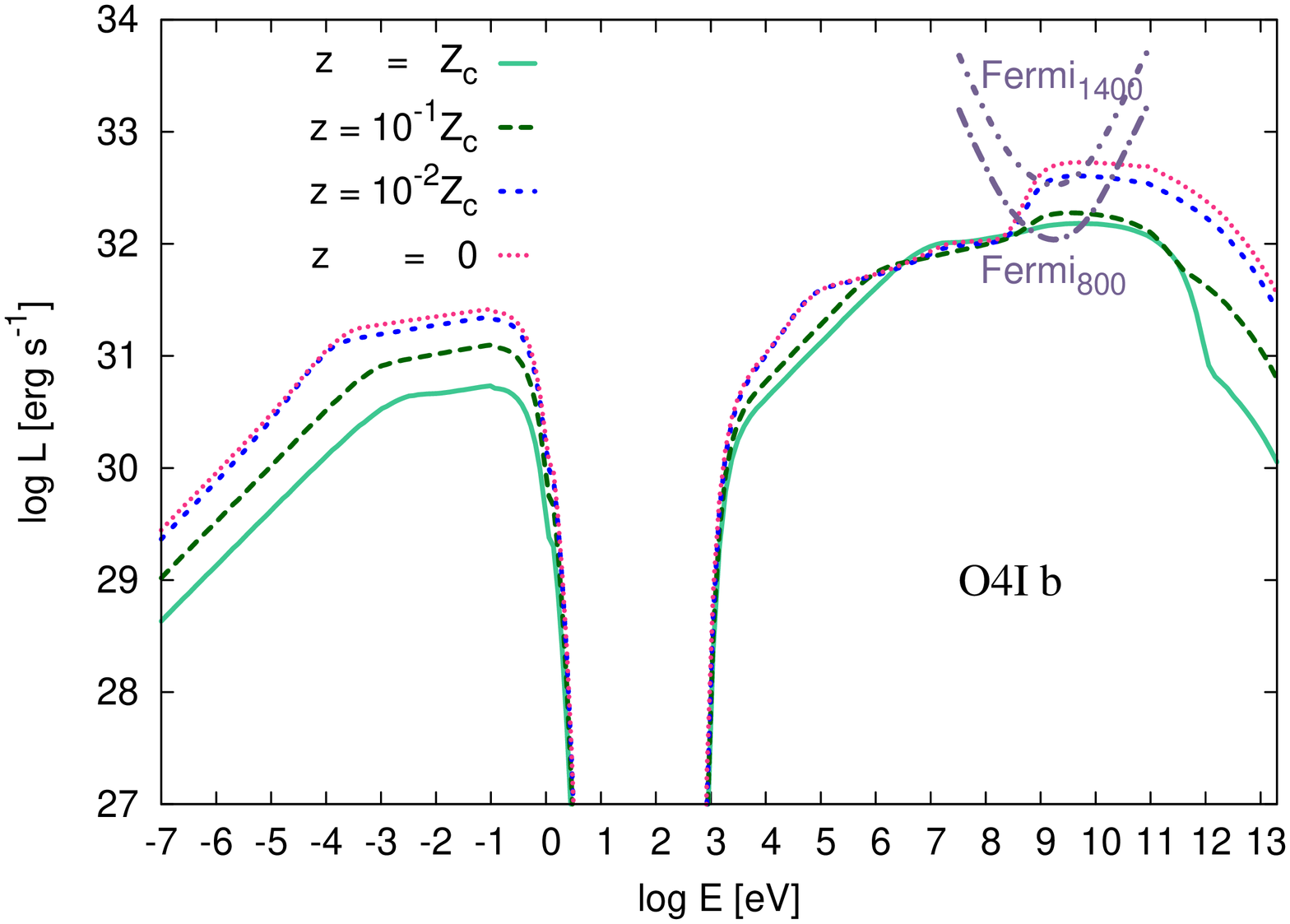}}
\caption{Total luminosity curves at $z$ = $0$, $10^{-2}Z_{\rm c}$, $10^{-1}Z_{\rm c}$, and $Z_{\rm c}$, for a O4I star, \emph{case a} (top) and \emph{case b} (bottom). {\it Fermi} sensitivities  at 0.8 and 1.4 kpc are also shown.}
\label{all-O4}
\end{center}
\end{figure}

\begin{figure}
\begin{center}
\resizebox{1.\columnwidth}{!}{\includegraphics[trim=0cm .8cm 0cm 0cm, clip=true,angle=270]{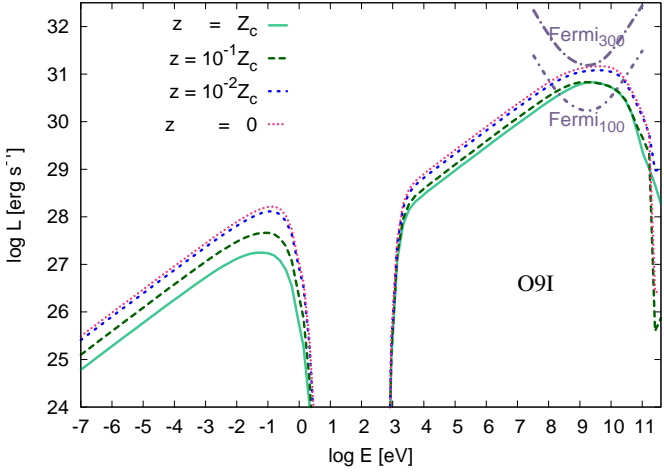}}
%\resizebox{1.\columnwidth}{!}{\includegraphics[trim=0cm .8cm 0cm 0cm, clip=true,angle=270]{all_O9_a100_abs_ii.eps}}
\caption{Total luminosity curves at $z$ = $0$, $10^{-2}Z_{\rm c}$, $10^{-1}Z_{\rm c}$, and $Z_{\rm c}$, for a O9I star. {\it Fermi} sensitivities  at $100$ pc  and 300 pc are also shown.}
\label{all-O9}
\end{center}
\end{figure}

\begin{figure}
\begin{center}
\resizebox{1.\hsize}{!}{\includegraphics[trim=0cm 0cm 0cm 0cm, clip=true,angle=270]{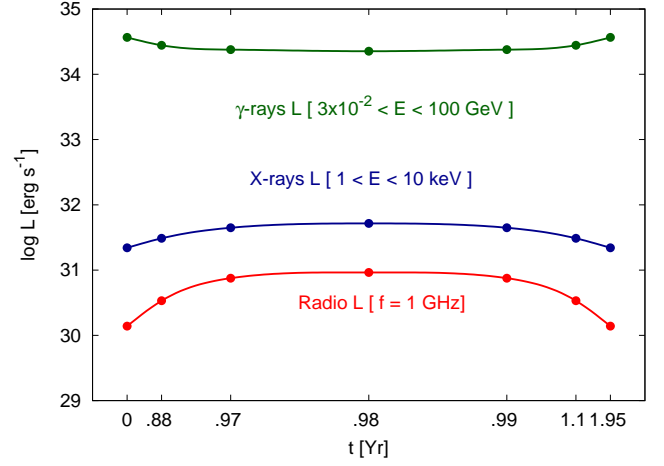}}
\resizebox{1.\hsize}{!}{\includegraphics[trim=0cm 0cm 0cm 0cm, clip=true,angle=270]{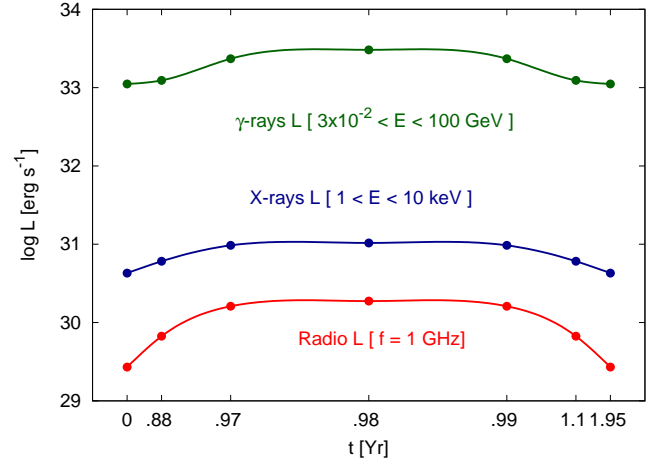}}
\caption{Variability curves at $z$ = $0$, $10^{-2}Z_{\rm c}$, $10^{-1}Z_{\rm c}$, and $Z_{\rm c}$, for a O4I star,  \emph{case a}  (top) and \emph{case b} (bottom).}
\label{vari-O4}
\end{center}
\end{figure}

\begin{figure}
\begin{center}
\resizebox{1.\hsize}{!}{\includegraphics[trim=0cm 0cm 0cm 0cm, clip=true,angle=270]{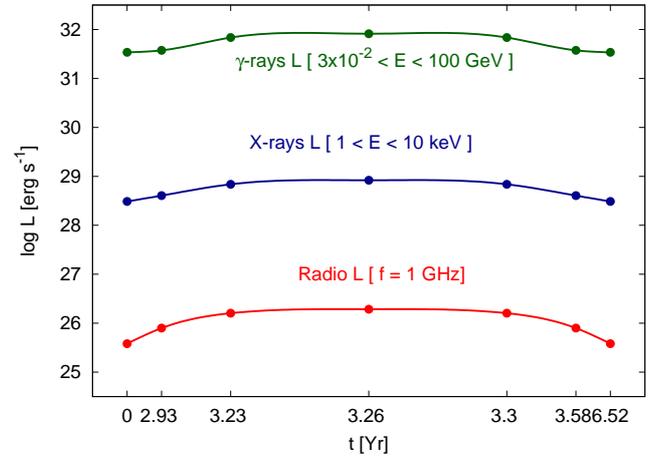}}
\caption{Variability curves at $z$ = $0$, $10^{-2}Z_{\rm c}$, $10^{-1}Z_{\rm c}$, and $Z_{\rm c}$, for a O9I star.}
\label{vari-O9}
\end{center}
\end{figure}

As the star moves through the density gradient, the emission varies with a timescale $\sim$ $Z_{\rm c}/V_{\star}$. For $Z_{\rm c}$ $\sim$ $10^{-4}$ pc, and the values adopted for $V_{\star}$, the variability timescale for the O4I star is $\sim$ 1 yr, and  $\sim$ 3 yr for the case of the O9I, with the parameters given in Table 1. Figures \ref{vari-O4} and \ref{vari-O9} show the integrated luminosity for the energy ranges of radio (1 GHz), X-rays ($1-10$ keV), and gamma-rays ($3 \times 10^{-2}-100$ GeV).

Nearby giant MCs are located at distances $d$ $\geq$ 100 pc. The O4I system,  for \emph{case a}, can be a variable gamma-ray source, detectable by {\it Fermi} at every $z$  for a wide range of distances. In \emph{case  b}, the detection can occur for $d$ $\leq$ 1 kpc pc (see Fig. \ref{all-O4}).  The O9I system might be detectable  at every $z$ for $d$ $\leq$ 300 pc (see Fig. \ref{all-O9}). In this case, the system would be a variable source, but with a longer variability timescale because the star moves slower. At larger distances the source
should not be detectable by {\it Fermi}, although it might appear as a weak source for the future Cherenkov Telescope Array (CTA, see Actis et al. 2011).

The gamma-ray and/or radio emission, in some cases, can be detectable only at the maximum of the light curve. Years later the emission might reappear as the stars travel through other  denser regions  in the MC. These sources might turn on and off within years. This situation might occur for the O4I system, in  \emph{case b}, when $d$ $\sim$ 1.4 kpc (see Fig.  \ref{all-O4}), and for the O9I system  when $d$ $\sim$ 300 pc (see Fig. \ref{all-O9}).

The weak or undetectable sources for {\it Fermi} might be detected by the future CTA, since it is expected to reach higher sensitivities  of almost one order of magnitude  better than {\it Fermi} at $E \sim 100$ GeV. In particular, CTA might be able to detect runaway O9I stars, which are more common in the solar neighbourhood.  

If the bowshock of a well-identified massive runaway star moving through an MC could be detected as a gamma-ray source, variations in its emission  will allow the study of the fine structure of the MC, which  is not possible at other energy bands where  the emission is highly absorbed.

The existence of a population of galactic variable gamma-ray sources is suspected since the epoch of the Energetic Gamma Ray Telescope (EGRET). Moreover, a statistically very significant positional correlation was found between gamma-ray sources in the third EGRET catalogue for OB star associations (e.g., Romero, Benaglia \& Torres 1999; Torres et al. 2001).

In the second {\it Fermi} catalogue (Nolan et al. 2012), 352 sources previously listed in the first catalogue (Abdo et al. 2010) did not show up. Most of these sources are concentrated
along the Galactic plane. Although some sources may have disappeared due to improvements in the model of the diffuse background, which is most intense at low latitudes, some other sources might intrinsically vary their fluxes from one catalogue to another.  This could be the case of massive runaway stars moving through molecular clouds.

We conclude that under some assumptions, bowshocks of massive runaway stars travelling through an MC might be variable gamma-ray sources; their time variability scale depends on the size scales of density inhomogeneities and the stellar velocities. Under some conditions, these sources might be galactic gamma-ray sources turning on and off over years. In this work, then, we propose a putative new class of galactic variable  gamma-ray source.

\begin{acknowledgements}
We thank an anonymous referee and Dr. M. Reynoso for valuable comments. This work is supported by PIP 0078 (CONICET) and PICT 2007-00848/2012-00878, Pr\'estamo BID (ANPCyT). 
\end{acknowledgements}

\end{document}